\documentstyle[12pt,graphicx]{article}
\title{Dark matter from $SU(4)$ model. }
\author{G.E. Volovik\\
Low Temperature Laboratory,
Helsinki University of Technology\\
P.O.Box 2200, FIN-02015 HUT, Finland\\
and\\
L.D. Landau Institute for Theoretical Physics,
  Moscow\\
}
\begin{document}
\maketitle
\begin{abstract}
{The left-right symmetric Pati-Salam model of the
unification of quarks and leptons is based on $SU(4)$ and $SU(2)\times SU(2)$
symmetry groups.  These groups are naturally extended to include the classification
of families of quarks and leptons. We assume that the family group (the group which
unites the families)  is also the $SU(4)$ group. The properties of
the fourth-generation of fermions are the same as that of the ordinary-matter
fermions in first three generations  except for the family charge of the
$SU(4)_F$ group:
$F=(1/3,1/3,1/3,-1)$, where $F=1/3$ for fermions of
ordinary matter and $F=-1$ for the fourth-generation fermions. The difference in
$F$ does not allow the mixing between ordinary and fourth-generation fermions.
Because of the conservation of the
$F$ charge, the creation of baryons and leptons in the process of electroweak
baryogenesis must be accompanied by the creation of fermions of the 4-th
generation. As a result the
the excess $n_B$ of baryons over antibaryons leads to the excess $n_{4\nu}=N-{\bar
N}$ of neutrinos over antineutrinos in the 4-th generation with $n_{4\nu}= n_B$.
This massive neutrino may form the non-baryonic dark matter. In
principle the mass density of the 4-th neutrino
$n_{4\nu} m_{N}$ in the Universe can  give the main contribution to the dark
matter, since the lower bound on the neutrino mass $m_N$ from the data on decay of
the
$Z$-bosons is
$m_N>m_Z/2$. The straightforward prediction of this model leads
to the amount of cold dark matter relative to baryons, which is an order of
magnitude bigger than allowed by observations. This inconsistency may be avoided
by non-conservation of the
$F$-charge.}
\end{abstract}

\section{Introduction}

Three-family structure of fermions is one of the most puzzling features of the
Standard Model. The other puzzles comes from the baryonic asymmetry of
the Universe and from astrophysical observations of the dark matter, whose
particles are most probably beyond the Standard Model. Here we make an attempt to
connect these three phenomena in the framework of the
$SU(4)$ scheme with four families of fermions. We discuss why the classification
scheme based on the $SU(4)$ groups is preferrable and how it leads to the
non-baryonic dark matter in the Universe if the Universe is baryonic
asymmetric. 

 Breaking of the  $SU(4)$ symmetry
 between four generations to its $SU(3)\times U(1)_F$ subgroup separates
the fermions of the fourth generation from the fermions of the other three
families. If the charge $F$ -- the generator of the $ U(1)_F$ group -- is conserved
in the process of baryogenesis, as it occurs for example in electroweak
baryogenesis, the formation of the baryon asymmetry automatically leads to
formation of dark matter made of the massive fourth-generation neutrinos. In such
scenario the mass density of the dark matter in the Universe essentially exceeds
the mass density of the baryonic matter.

\section{Pati-Salam model}

In the current Standard Model three families 
(generations) of fermions have identical properties above the electroweak energy
scale. Each family contains 16 Weyl fermions (8 left and 8 right) which transform
under the gauge group
${\rm G}(213)= SU(2)_{L}\times U(1)_Y\times SU(3)_C$ of weak, hypercharge and 
strong interactions respectively. In addition they are characterized by two global
charges: baryonic number $B$ and leptonic number $L$.

The Grand
Unification Theories (GUT) unify
weak, hypercharge and strong interactions into one big
symmetry group, such as
$SO(10)$, with a single coupling constant. 
There is, however, another group, the minimal subgroup of $SO(10)$,
which also elegantly unites 16 fermions in each generation, a type of Pati--Salam
model
 \cite{PatiSalam,PatiSalam2,Foot}
with the symmetry group
${\rm G}(224)=SU(2)_{L}\times SU(2)_{R} \times SU(4)_C$. This left-right
symmetric group preserves all the important properties of $SO(10)$,
but it has advantages when compared to the
$SO(10)$ group \cite{PatiNew,Pati2002,Adler2}. In particular, the electric
charge is left--right symmetric:
\begin{equation}
Q=\frac{1}{ 2} (B-L) +T_{R3} +T_{L3} ~,
\label{SymmetricElectricCharge}
\end{equation}
where $B-L=( \frac{1}{ 3},   \frac{1}{ 3},  \frac{1}{ 3}, -1)$ is the
difference between baryon and lepton numbers (the generator of $SU(4)_C$ color
group), and
$T_{R3}$ and
$T_{L3}$ are left and right isotopic spins (generators of the groups $SU(2)_{R}$
and 
$SU(2)_{L})$.  Also the ${\rm G}(224)$ group nicely fits the classification scheme
of quantum vacua based on the momentum-space topology of fermionic propagators
\cite{Book}.

\begin{figure}
  \centerline{\includegraphics[width=0.8\linewidth]{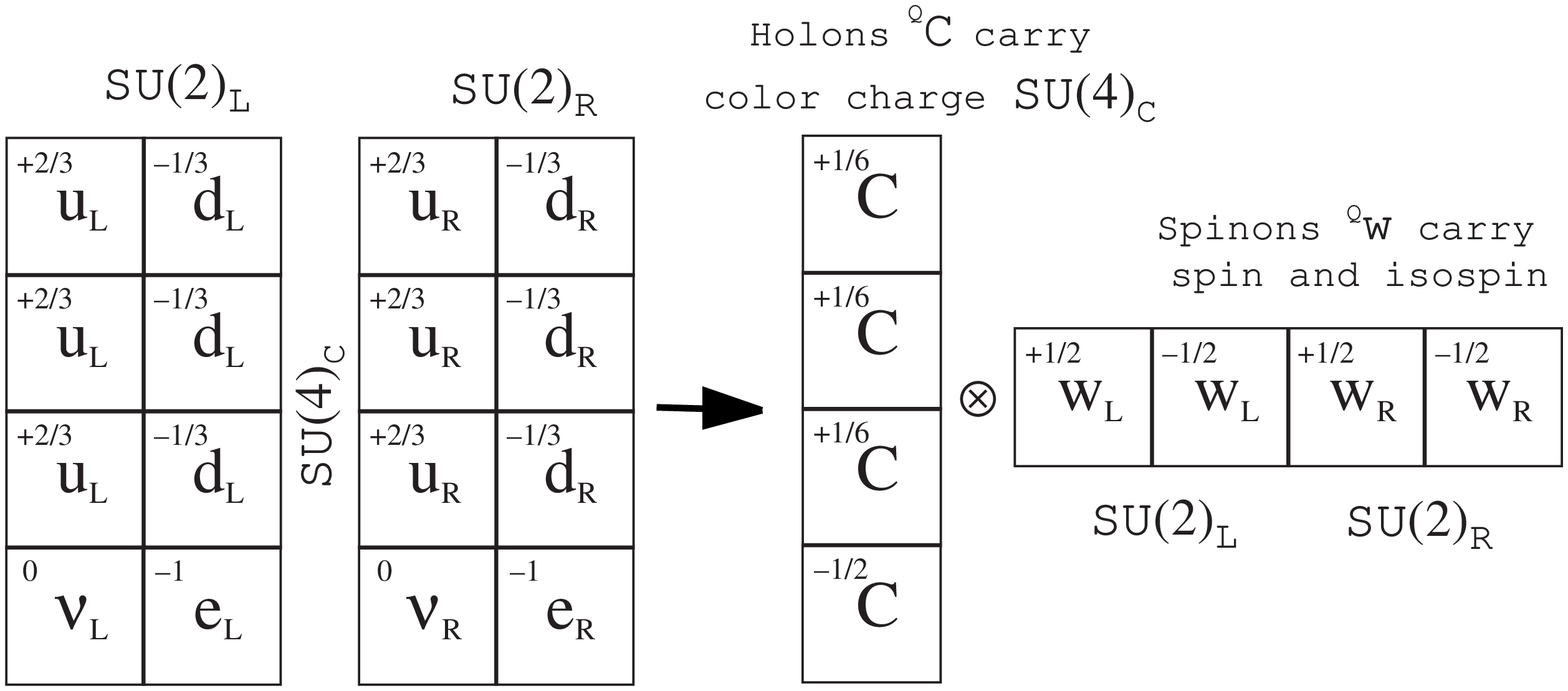}}
  \caption{Terazawa scheme of composite fermions $wC$ as bound states of
$w$-fermions and $C$-bosons. Numbers show the electric charge.}
  \label{FirstFamilyFig}
\end{figure}

According to Terazawa \cite{Terazawa} the 16 fermions of each generation can be
represented as the product
$Cw$ of 4 bosons and 4 fermions in Fig. \ref{FirstFamilyFig}.
The Terazawa scheme is similar to the slave-boson approach in condensed matter,
where the spinons are fermions which carry spin and holons are bosons which carry
electric charge  \cite{Book}.  Here the "holons" $C$ form the color
$SU(4)_C$ quartet of spin-0
$SU(2)$-singlet particles with $B-L$ charges of the $SU(4)_C$ group  and electric
charges $Q=( \frac{1}{6},  \frac{1}{6}, \frac{1}{6}, -\frac{1}{2})$. The "spinons"
are spin-$\frac{1}{2}$ particles
$w$, which are color
$SU(4)_C$ singlets and $SU(2)$-isodoublets. 

\section{Family charge and fourth generation}

\begin{figure}
  \centerline{\includegraphics[width=0.6\linewidth]{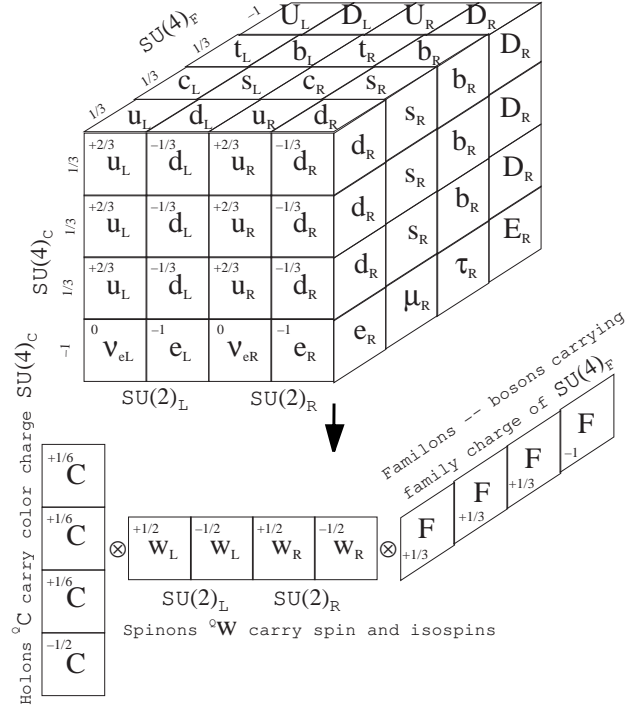}}
  \caption{Extended Terazawa scheme of composite fermions $wCF$ as bound states  of
$w$-fermions and $C$- and $F$-bosons.  }
  \label{FamiliesFig}
\end{figure}

What is missing in this scheme is fermionic families. The natural extension, which
also uses the $SU(4)$ symmetry, is to introduce the group
$SU(4)_F$ in the family direction (Fig. \ref{FamiliesFig}). This requires the
introduction of additional 4-th generation of fermions. Then the family charge $F$
comes as one of the generators of
$SU(4)_F$ in the same way as the $B-L$ charge comes as one of the generators of
$SU(4)_C$ color group: 
$F=( \frac{1}{3},  \frac{1}{3}, \frac{1}{3}, -1)$. Here $F=\frac{1}{3}$ is
assigned to ordinary matter in three generations, while
$F=-1$ corresponds to the fourth generation.  

The total number 64 of Weyl
fermions, each being described by 2 complex or 4 real functions, satisifies the
$2^n$ rule, which probably reflects the importance of the Clifford algebra in the
undelying physics \cite{WilczekZee}. This is another advantage of classification
based on $SU(2)$ and $SU(4)$ groups. This extension also fits the Terazawa scheme
which is modified by addition of extra boson. All  64 fermions  are now
represented as the product
$FCw$ in Fig. \ref{FamiliesFig}, where $F$ is the quartet of bosons with $F$
charges of the $SU(4)_F$.

Let us assume that the breaking of $SU(4)_F$ symmetry occurred in the same way
as the breaking of the color group: $SU(4)_C\rightarrow SU(3)_C\times U(1)_{B-L}$
and $SU(4)_F\rightarrow SU(3)_F\times U(1)_{F}$, where $U(1)_F$ is the group
generated by charge $F$. Then all the charges of
the fourth-generation fermions are the same as that of the ordinary-matter
fermions in first three generations  except for the family charge $F$. 
 This separates out 
the 4-th generation in the same manner as leptons -- the fourth color -- are
separated from quarks. The 4-th generation fermions cannot mix with ordinary
fermions in the first three generations in the same manner as leptons cannot mix
with quarks. This is the main difference from the democratic scheme of sequential
generations, which assumes that all generations are equal. The important
consequence of such a discrimination between the generations is that the excess of
the baryonic charge in the Universe must be accompanied by the uncompensated
neutrinos of the 4-th generation. 

\section{Dark matter and electroweak baryogenesis}

Let us assume that the $F$ charge is strictly conserved. This certainly occurs
in the process of electroweak baryogenesis, in which the baryons and leptons are
created due to the anomalous non-conservation of the baryonic and leptonic
numbers, while the chiral anomaly conserves the charges
$F$ and
$B-L$ (this does not depend on whether $SU(4)_F$ is a local or global group). The
conservation of these two charges gives
\begin{equation}
F= \frac{1}{3}(n_q+n_l)-(n_{4q} +n_{4l})=0
~,~B-L=\frac{1}{3}(n_q+n_{4q})-(n_l+n_{4l})=0~, 
\label{FCharge}
\end{equation}
 where $n_q$  and $n_l$ are (algebraic) densities of quarks and leptons  belonging
to the ordinary matter in three generations;  and 
$n_{4q}$ and $n_{4l}$ are that of the fourth generation. From Eqs.(\ref{FCharge})
one obtains the densities of the fourth-generation quarks and leptons in terms of
the ordinary matter (which mainly belongs to the first generation):
\begin{equation}
n_{4q}=n_l~,~n_{4l}=\frac{1}{3}n_q-\frac{2}{3}n_l~. 
\label{FermionNumber}
\end{equation}
If in our Universe $n_l=0$, i.e. the leptonic charge of electrons is compensated
by that of antineutrinos, one obtains
\begin{equation}
n_{4q}=0~,~n_{4l}=\frac{1}{3}n_q=n_B~, 
\label{FermionNumber2}
\end{equation}
where $n_B$ is the algebraic number density of baryons in our Universe.

This means that any excess of the baryonic charge in the Universe is accompanied by
the excess $n_{4\nu}=N-{\bar N}=n_B$ of the neutrinos over antineutrinos in the
fourth generation (we assume that the mass of the 4-th
generation neutrino is smaller than that of the 4-th
generation electron,
$m_N<m_E$). If this is correct, the 4-th generation  neutrinos adds to the mass
of our Universe and may play the role of the dark matter. The mass $m_N$ of the
4-th generation (dark) neutrino can be expressed in terms of the mass of baryon
(nucleon) $m_B$ and the energies stored in the baryonic matter (with fraction
$\Omega_B$ of the total mass in the flat Universe) and non-baryonic dark matter
(with fraction $\Omega_{DM}$):
\begin{equation}
\frac {m_{N}}{m_B}= \frac {\Omega_{DM}}{ \Omega_B} ~.
\label{MassDarkNeutrino}
\end{equation}

Masses of the 4-th generation of fermions must be below the electroweak energy
scale, since the masses of all the fermions (ordinary and of 4-th generation) are
formed due to violation of the electroweak symmetry, as follows from the
topological criterion of mass protection (see Chapter 12 of Ref. 
\cite{Book}). There is also the lower constraint on the 4-th neutrino mass which
comes from the measured decay properties of the
$Z$-boson:
$m_{N}>m_Z/2=45.6$ GeV (see e.g.
\cite{DubickiFroggatt} and somewhat stronger constraint $m_{N}> 46.7$ GeV from
the $Z$-resonance lineshape in
\cite{Bulanov}). Then the fourth generation contribution to the dark matter must be
\begin{equation}
 \Omega_{DM}= \Omega_B \frac{m_{N} }{ m_B}> 50 \Omega_B~.
\label{MassDarkNeutrino}
\end{equation}
Since in our spatially flat Universe 
$\Omega_{DM}< 1$,  this gives the constraint on the baryonic mass in the
Universe: $\Omega_B< 0.02$. 
On the other hand, since the baryonic density in the luminous matter is 
$\Omega_B\sim 0.004$, the constraint on the dark matter mass is  $\Omega_{DM}>0.2$.

There are, however, the other constraints on the mass of $m_{N}$ of heavy neutrino
obtained from particle and astroparticle implications of the 4-th generation
neutrinos
\cite{Khlopov}, from which it follows that $m_{N}$ must be larger than 200 GeV.
For the early limits on heavy neutrino contribution to the density of
dark matter, which follow from consideration of neutrino-antineutrino
annihilation, see refs \cite{Vysotsky}.
On the other hand the electroweak data fit prefers the 50 GeV neutrinos
\cite{Okun}, though the masses as high as 200 GeV are not excluded. 

One must also take into account that the dark matter contribution  $\Omega_{DM}$ is
only a fraction of unity, as follows from observational data on CMB angular
temperature fluctuations \cite{Spergel}. On the other hand the baryonic content of
dark matter is about 10 times bigger than the baryon density in the visible
matter, as follows independently both from the structure of acoustic peaks in the
angular power spectrum of CMB temperature fluctuations  \cite{Spergel}
and from the primordial nucleosynthesis theory plus
observed abundance of light elements ($^4$He, D, $^3$He and $^7$Li) 
\cite{PrimordialNucleosynthesis}. As a result one finds that $\Omega_{DM}/\Omega_B
< 7$. Together with the constraint on  $m_{N}$ this indicates that the excess
$n_{4\nu}$ of neutrinos over antineutrinos must be smaller (maybe by one or two
orders of magnitude) than $n_B$. This would mean that the $F$-charge is not
strictly conserved.

\section{Conclusion}

It appears that the classification scheme based on the $SU(2)$ and $SU(4)$ groups
has many interesting features. Here we considered one of them -- the discrimination
between the ordinary-matter fermions and the fourth
generation. If the family charge $F$ is strictly conserved, the fermions of the
fourth-generation neutrinos with density $n_{4\nu}=n_B$
 must be necessarily present in the baryonic asymmetric
Universe to compensate the family charge. Moreover, it is not excluded
that it is the asymmetry in the fourth-generation neutrinos which is primary
and it serves as a source of the baryonic asymmetry of the Universe.  

These massive
neutrinos form the stable dark matter with mass density exceeding the baryonic
mass density by the ratio of masses of neutrino and baryon $m_{N}/m_B$.
According to the well known physics of the decay of
$Z$-boson, this factor must be larger than 50. Whether this
is an acceptable solution to the dark matter problem depends on details of the
${\rm G}(2244)= SU(2)_{L}\times  SU(2)_{R}\times SU(4)_C\times SU(4)_F$ model. The
above relation between the baryonic and non-baryonic masses of the Universe is
based on the assumption of the conservation of the $F$-charge.   If this
requirement is weakened, the dark matter density coming from the fourth generation
of fermions will depend on the details of the baryo- or neutrino-genesis and on the
further cosmological evolution of ordinary and dark matter.

 I am grateful to S.L. Adler for e-mail
correspondence, J.C. Pati for encouraging comment,   M.Yu. Khlopov and T.
Vachaspati for numerous discussions, and  L.B. Okun and A.A. Starobinsky for
careful reading of the manuscript and valuable suggestions.
 This work was
supported by
ESF COSLAB Programme and by the Russian Foundations for
Fundamental Research.

\end{document}